\documentclass[a4paper]{jpconf}
\usepackage{graphicx}
 \usepackage{wrapfig}

\newcommand{\muB}{$\mu_{\rm B}$}
\newcommand{\sNN}{$\sqrt {{s_{\rm NN}}}$}

\newcommand{\MV}{$\sigma^{2}/M$}
\newcommand{\Ss}{$S\sigma$}
\newcommand{\KV}{$\kappa\sigma^{2}$}
\begin{document}

\bigskip
%\textbf{WWND 2014 Proceedings}

\title{Recent results on event-by-event fluctuations from the RHIC
  Beam Energy Scan program in the STAR experiment}

\author{Nihar Ranjan Sahoo (for the STAR Collaboration)}

\address{Texas A\&M University, College Station, Texas 77843, USA}

\ead{nihar@rcf.rhic.bnl.gov}

\begin{abstract}
Event-by-event fluctuations of global observables in relativistic
heavy-ion collisions are studied as probes
for the QCD phase transition and as tools to search for critical phenomena near
the phase boundary. 
Dynamical fluctuations in mean
transverse momentum, identified particle ratios and conserved
quantities (such as net-charge, net-baryon) are expected to 
provide signatures of a de-confined state of matter. Non-monotonic
behavior in the higher-moments of conserved quantities 
as a function of beam energy and collision centrality are proposed
as signatures of the QCD critical point. To study the QCD phase transition and locate the critical
point, the STAR experiment at RHIC has collected a
large amount of data for Au+Au collisions
from $\sqrt{s_{NN}} = 7.7-200$~GeV in the RHIC Beam Energy Scan (BES) program.
We present the recent beam energy scan results on dynamical
fluctuations of particle ratios and two-particle
 transverse momentum correlations at mid-rapidity.
Higher-moments of the net-charge and net-proton  multiplicity distributions as a function of beam energy will be
presented. We give a summary of what has been learnt so far and future
prospectives for the BES-II program.
\end{abstract}

\section{Introduction} 

Importance of the Event-by-Event (E-by-E) fluctuations in
heavy-ion collisions has been studied and discussed in a series of
papers~\cite{Heiselberg,Byam,Stodolsky,Stephanov-Rajagopal-Shuryak}.
The E-by-E fluctuation observables could provide information about the QCD
phase transition and a possible signature of QCD critical point. RHIC, at Brookhaven National
Laboratory, started the Beam Energy Scan (BES) program since 2010 in order to
explore the QCD phase diagram. The STAR
experiment, at RHIC, has collected data at \sNN~= 39 GeV down to 7.7 GeV for this BES program.

\section{RHIC Beam Energy Scan Program}

 The main goals of the BES program are to (1)
 locate the existence of the QCD critical point, (2) find evidence of the
 first order phase transition in the QCD Phase diagram, and (3)
 understand the properties of the QGP as a function of  \muB. The
 variation of RHIC colliding energy from \sNN~= 7.7 to 39 GeV,
 along with 62.4 and 200 GeV, covers the baryonic chemical potential
 from 410 to 20 MeV~\cite{Cleymans}. The RHIC beam energy scan program started in
 the year 2010 and the first phase of the beam energy scan has
 been finished in year 2014 with \sNN~= 14.5 GeV data taking. Details about
 the BES program are listed in Table~\ref{BES_table}.

The STAR experiment is able to measure various E-by-E observables to
 study the signatures of QGP-to-Hadron-Gas phase transition. Full 2$\pi$ azimuthal coverage, enriched
 particle identification capabilities and large uniform acceptance of STAR provides a
 suitable environment for the E-by-E characterizations of heavy-ion-collisions.
In order to achieve the goals of the BES program, the STAR collaboration have analyzed various E-by-E fluctuation measures. For the search of
QCD critical point, higher-moments of the conserved charge, like
net-charge, net-protons (proxy for net-baryon) and net-kaon (proxy for
net-strangeness) distributions, particle ratio fluctuation, and transverse momentum
fluctuations have been analyzed. Dynamical charge
fluctuations for the signature of QGP have been studied as well. An elaborate
discussion can be found in the subsequent sections.
\begin{table}[h]
\begin{center}
\begin{tabular}{|c |c |c |c |c |}
\hline
$\sqrt{s_{\rm NN}}$(GeV)   & \muB & Year &  Events
($\times10^{6}$) & Beam times (weeks)   \\
\hline
39       &  115   & 2010  & 130 & 2.0   \\
%\hline
27       &  155   & 2011  & 70 & 1.0   \\
%\hline
19.6       &  205    & 2011  & 36 & 1.5   \\
%\hline
14.5       &  206    & 2014  & 20 & 3.0   \\
%\hline
11.5       &  315    & 2010  & 12 & 2.0   \\
%\hline
7.7       &  420    & 2010  & 4 & 4.0   \\
\hline
\end{tabular}
\caption{Beam Energy Scan program details. The \muB~values are taken from Ref.~\cite{Cleymans}}
\label{BES_table}
\end{center}
\end{table}
 
\section{Search for QCD Critical Point }

Lattice QCD calculations reveal that
at vanishing $\mu_{B}$, the transition from QGP to hadron gas is a simple 
crossover~\cite{aoki}, whereas at large $\mu_{B}$, the phase
transition is of first order~\cite{aoki,ejiri,bowman,stephanov,fodor,gavai,cheng,indication}.
Therefore, one expects the existence of a critical point at the end of the
first order phase transition. Search for the QCD critical point has been one of
the major thrusts of the RHIC BES program. 

At the critical point, thermodynamic susceptibilities and the
correlation length ($\xi$) of the system are expected to diverge for large
samples in equilibrium. The finite volume effect puts a constraint on the
divergence of the thermodynamic susceptibilities and $\xi$ of the
system. The phenomenon of critical slowing down in the vicinity of the critical point drives the system away
from thermodynamic equilibrium, so $\xi$ reaches a maximum value of
around $1.5-3$~fm~\cite{berdnikov,athanasiou}. Besides these
challenges, in heavy-ion-collision experiments, the signature of
critical point could be observed if the critical point is close enough to the freeze-out curve~\cite{SG_PLB}.  Various observables such as, higher-moments of the
conserved charge,  transverse momentum fluctuations, particle ratio
fluctuations, have been proposed in order to
search for the QCD critical point. Detailed discussions about these
analysis and their observables can be found in the following sections.

\subsection{{\bf Higher-moments of  net-charge and net-proton distributions}}
Moments of the conserved charge distributions (such as the mean ($M$), standard
deviation ($\sigma$), skewness ($S$) and kurtosis ($\kappa$)) have
been proposed as important observables for the signature of the QCD critical point in the current wisdom. The higher-moments of conserved charge distributions are related to the
respective higher order thermodynamical susceptibilities and also $\xi$
of the system~\cite{athanasiou,Stephanov_corrl}. In order to cancel the undetermined volume term in
the higher-moments, ratio (or product) like \MV, \Ss, and \KV~are used. The signature of non-monotonicity of these observables
is expected if there is a nearby critical point in QCD phase transition.
Recently, the STAR experiment reported net-charge~\cite{star_net_charge} and
net-proton~\cite{star_net_proton} results from the BES program. 

In an analysis of higher-moments, various sophisticated techniques, like
the finite bin width effect~\cite{CBW}, finite efficiency
correction~\cite{Bzdak} for the higher-moments, have been incorporated. In order to get
precise statistical uncertainty, different statistical error estimation
methods (like the Delta Theorem, Bootstrap method, etc.) have been
throughly studied and utilized. On the other hand, new centrality definitions (like
uncorrected charge multiplicity within $0.5<|\eta|<1.0$ for net-charge
analysis and that of other than identified protons and antiprotons
within pseudo-rapidity $|\eta| < 1.0$ for net-proton) have also
been studied and implemented. 

\begin{figure}
  \begin{center}
    \includegraphics[width=0.54\textwidth]{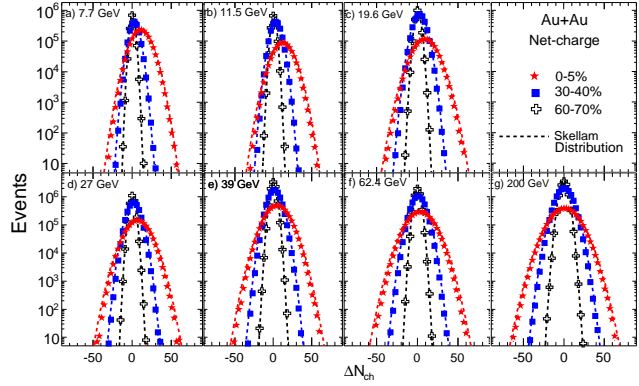}
  \end{center}
  \caption{ The net-charge distributions, for Au+Au collisions at 7.7 to 200 GeV within $|\eta| < 0.5$ and $0.2 < p_{T} < 2.0$ GeV/c, drawn with the Skellam distribution for three different centralities ($0-5\%, 30-40\%, 60-70\%$).}
\label{netQ_dist}
\end{figure}

The positive ($N_{+}$) and negative ($N_{-}$)
charged particle multiplicities are counted within  $|\eta|<0.5$ and 
$0.2 < p_{\rm{T}} < 2.0$~GeV/{\it{c}} 
(after removing protons and anti-protons with $ p_{\rm{T}} <
400$~MeV/{\it c}) to calculate net-charge ($ N_{+} - N_{-}$)
in each event. The protons ($N_{p}$) and anti-protons ($N_{\bar{p}}$)
are counted at mid-rapidity ($|y| < 0.5$) in the range $0.4 < p_{T} < 0.8$
GeV/{\it c}. In addition to these, various track selection cuts have been
used. Details about these cuts can be found elsewhere in
Ref~\cite{star_net_proton,star_net_charge}. Figure~\ref{netQ_dist}
shows the net-charge multiplicity distribution for BES energies along
with the Skellam distribution (the difference of two Poisson
distributions assuming $N_{+}$ and $N_{-}$ are independent Poisson
distributions)~\cite{STAR_netQ_prelim}. The Skellam distribution qualitatively
follows the net-charge distributions for all energies. Similarly, the
net-proton distributions can be found in Ref.~\cite{star_net_proton}.

\begin{figure}
\begin{center}
\includegraphics[width=0.40\textwidth]{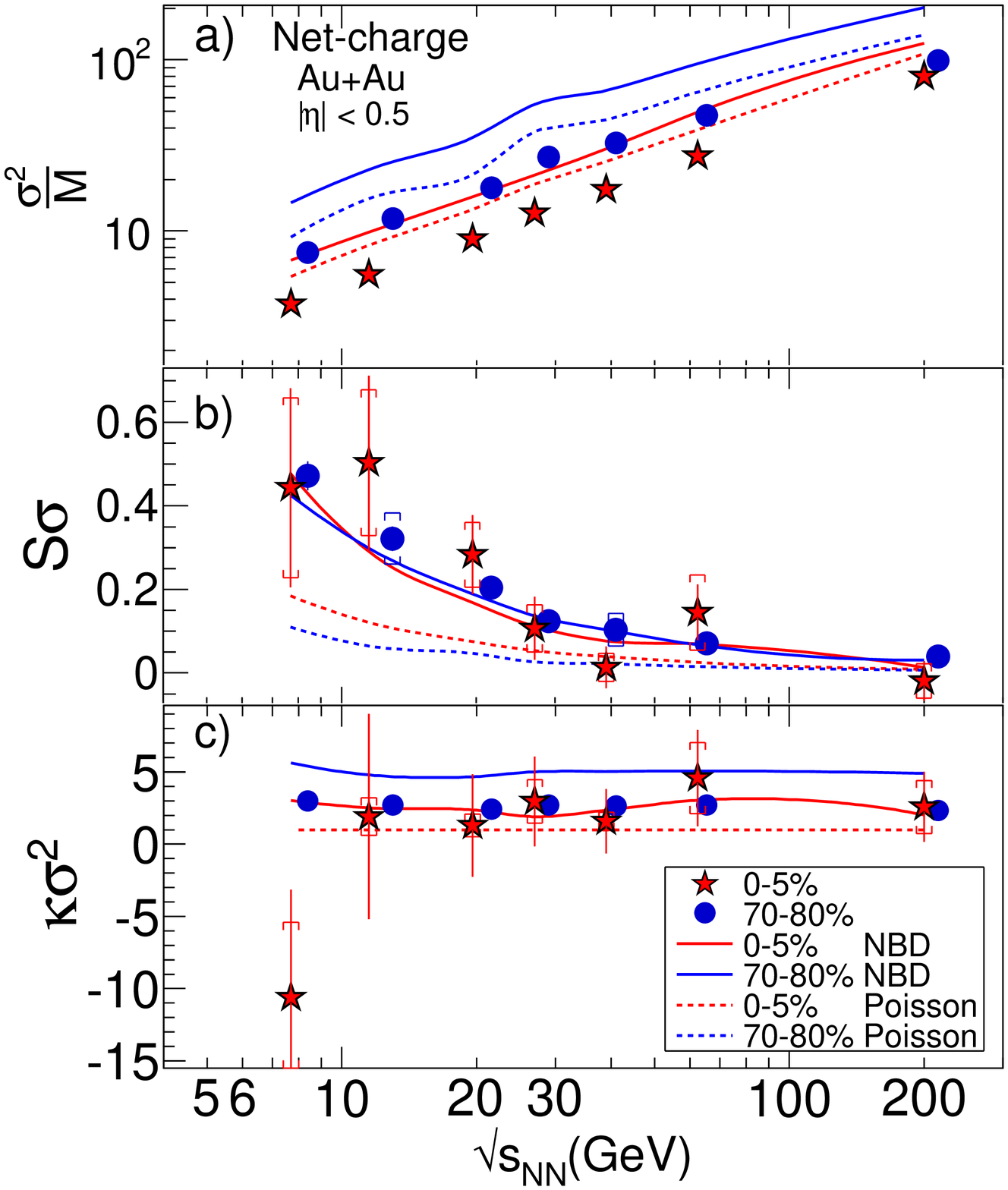}
\includegraphics[width=0.41\textwidth]{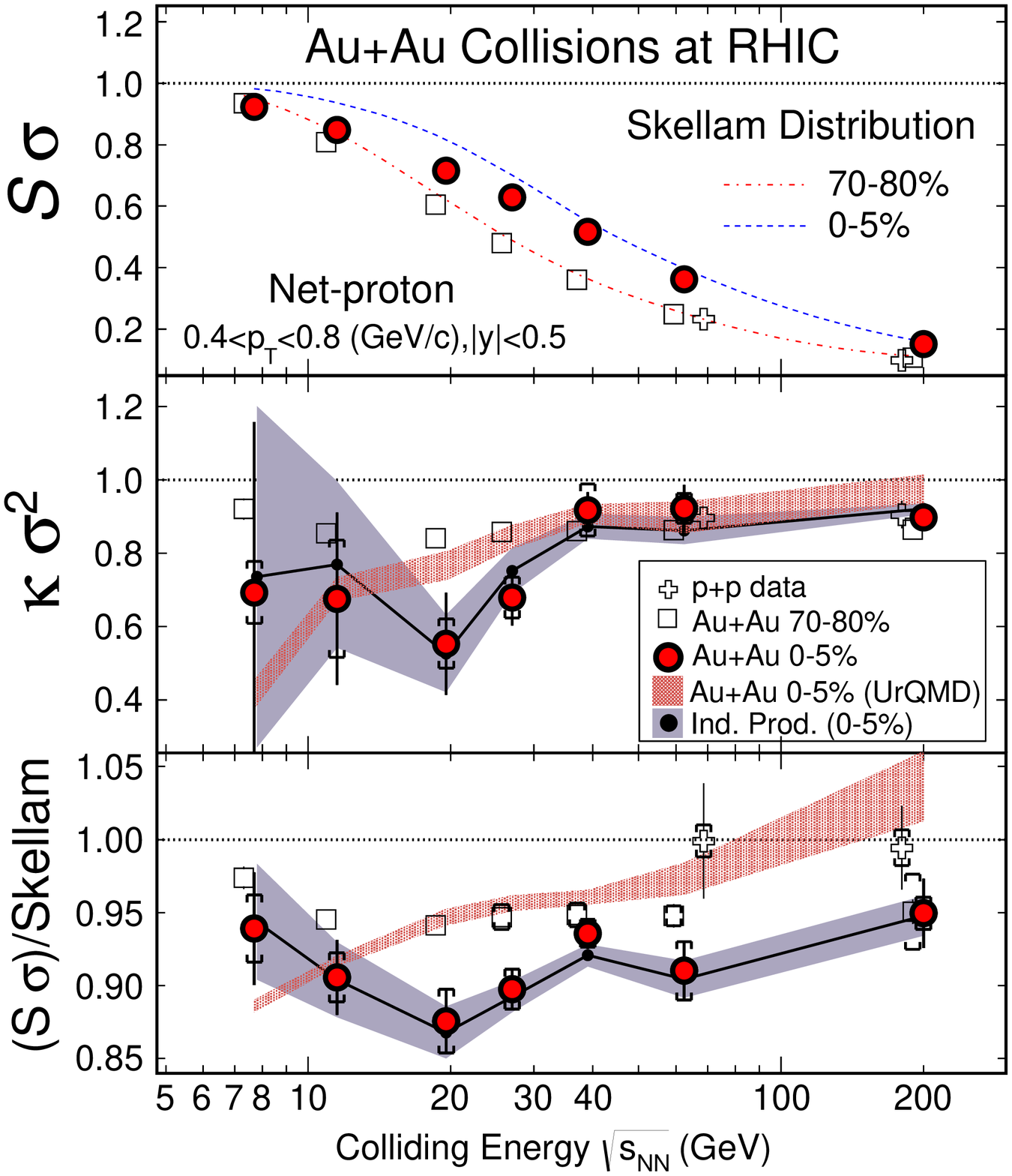}
\end{center}
\caption{\label{HM_BES}(left panel) Beam-energy dependence of (a) \MV, (b)
 \Ss, and (c) \KV, after all corrections, for most central (0-5$\%$) and peripheral (70-80$\%$) bins~\cite{star_net_charge}. The error bars are
statistical and the caps represent systematic errors. Results from the
Poisson and the NBD baselines are superimposed. 
The values of \KV~for Poisson baseline are always unity. (Right panel)
Collision energy and centrality dependence of the net proton \Ss~and \KV~from Au+Au and p+p collisions at RHIC~\cite{star_net_proton}. 
Skellam distributions (dash lines in top panel), UrQMD (red shaded
band) and indepedent particle production assumption expectation
(shaded solid band) for corresponding collision centralities are
shown.}
\end{figure}

Reconstruction efficiency corrected values for the products of higher-moments of net-charge and net-proton number distributions
have been shown in Fig.~\ref{HM_BES}. In order to understand these
results, different baseline measurements like Poisson and Negative
Binomial Distribution (NBD) baseline for
the net-charge higher-moments analysis are studied. Similarly, Poisson, Independent particle production
expectation and UrQMD model simulation have been performed for the net-proton
analysis. These baselines lack any QCD critical point phenomena, therefore it
is expected that any deviation from these baselines could be due to the
possible presence of new physics in the data. In general, both NBD and Poisson baseline calculations overestimate
the net-charge data. For central collisions, within the statistical
and systematic errors, the values \KV~of the net-charge
distribution at all energies are consistent with each other except for
\sNN~= 7.7 GeV. The significance of deviation of data with respect to
both the baselines remains within 2$\sigma$ in case of \Ss~and \KV. This
implies that the products of moments of net-charge distributions do
not show non-monotonic behavior as a function of beam energy within
statistical errors. The large statistical uncertainty plays an important
role for the net-charge higher-moments analysis and also puts a
constraint for any definite conclusion.  

On the other hand, the higher-moments of net-proton distribution results 
show the significance of deviations w.r.t Poisson
baseline and Hadron Resonance Gas model, for 19.6 and 27 GeV, with 3.2 and 3.4 for \KV, and 4.5 and
5.6 for \Ss~at 0-5$\%$ centrality, respectively. In order to
understand the effect of baryon number conservation effect on higher
moments of net-proton distributions, UrQMD model simulation for 0-5$\%$
centrality have also been compared. It shows that \KV~ and
\Ss/(Skellam), for UrQMD model simulations, monotonic decrease with
decreasing beam energy. Independent particle production
approach (by considering independent production of protons and
anti-protons in a given ensemble of events) have also been studied by
constructing net-proton cumulants using the expression $C_{n}(N_{p} -
N_{\bar{p}}) = C_{n} (N_{p}) + (-1)^{n}C_{n}(N_{\bar{p}})$, where
$C_{n} (N_{p})$ and $C_{n}(N_{\bar{p}})$ (n = 1, 2, 3 and 4) are the independently
calculated cumulants of protons and anti-protons, respectively. The
results based on independent particle production follow the data at
all energies implying a weak correlation strength between protons and
anti-protons in this analysis acceptance. 

 Large statistics and more energy points are needed to pin-point the exact position of the
deviation from baseline as a function of beam energy. On the other hand, various
final state effects like diffusion of conserved charge fluctuation, the effect of resonance decay
within a given rapidity window, the effect of volume fluctuations, finite time and volume effects may play significant role
in this observable. Large collection of events is needed, and systematic studies both on the experimental and
theoretical sides need to be explored. 

\begin{figure}
\begin{center}
\includegraphics[width=0.40\textwidth]{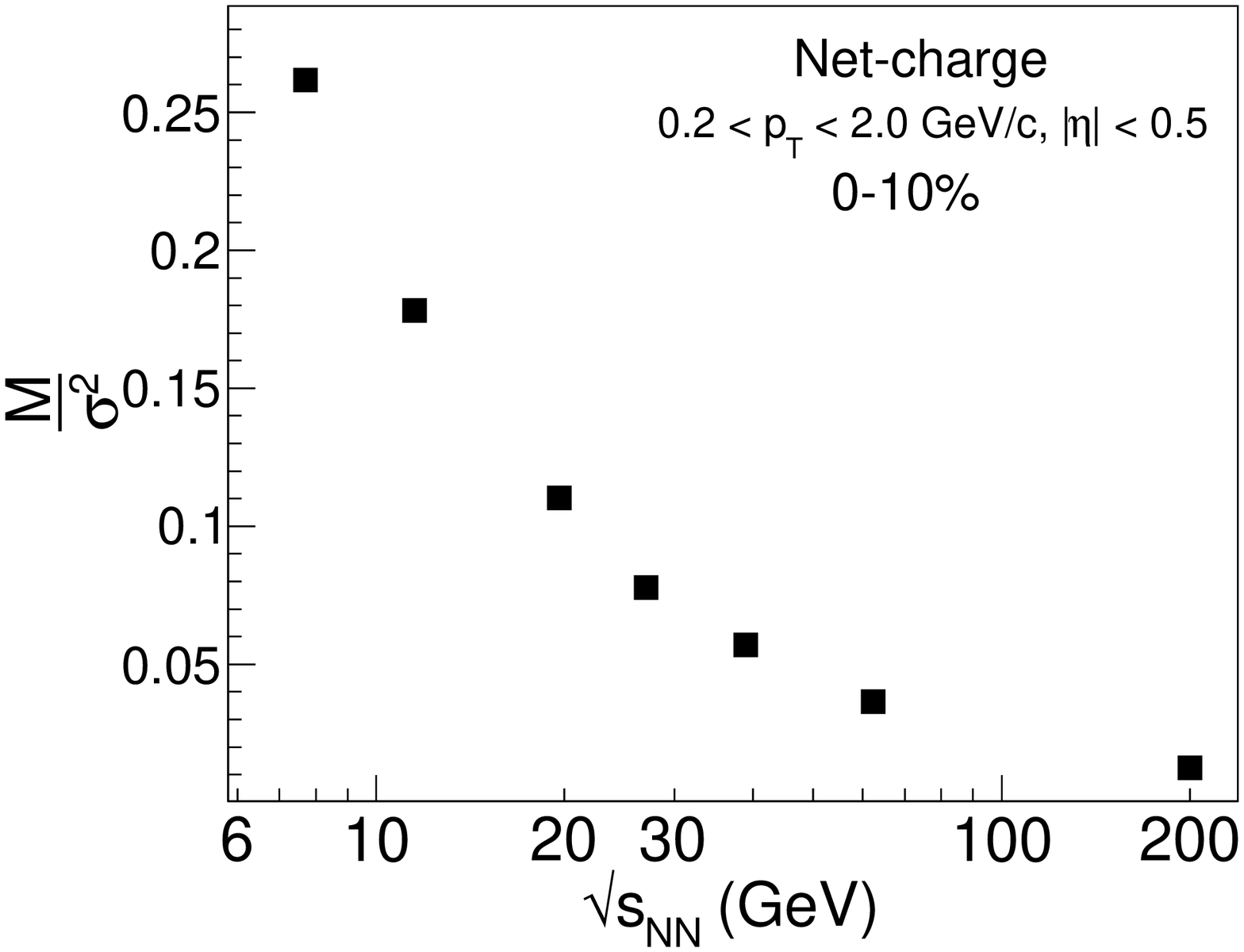}
\includegraphics[width=0.41\textwidth]{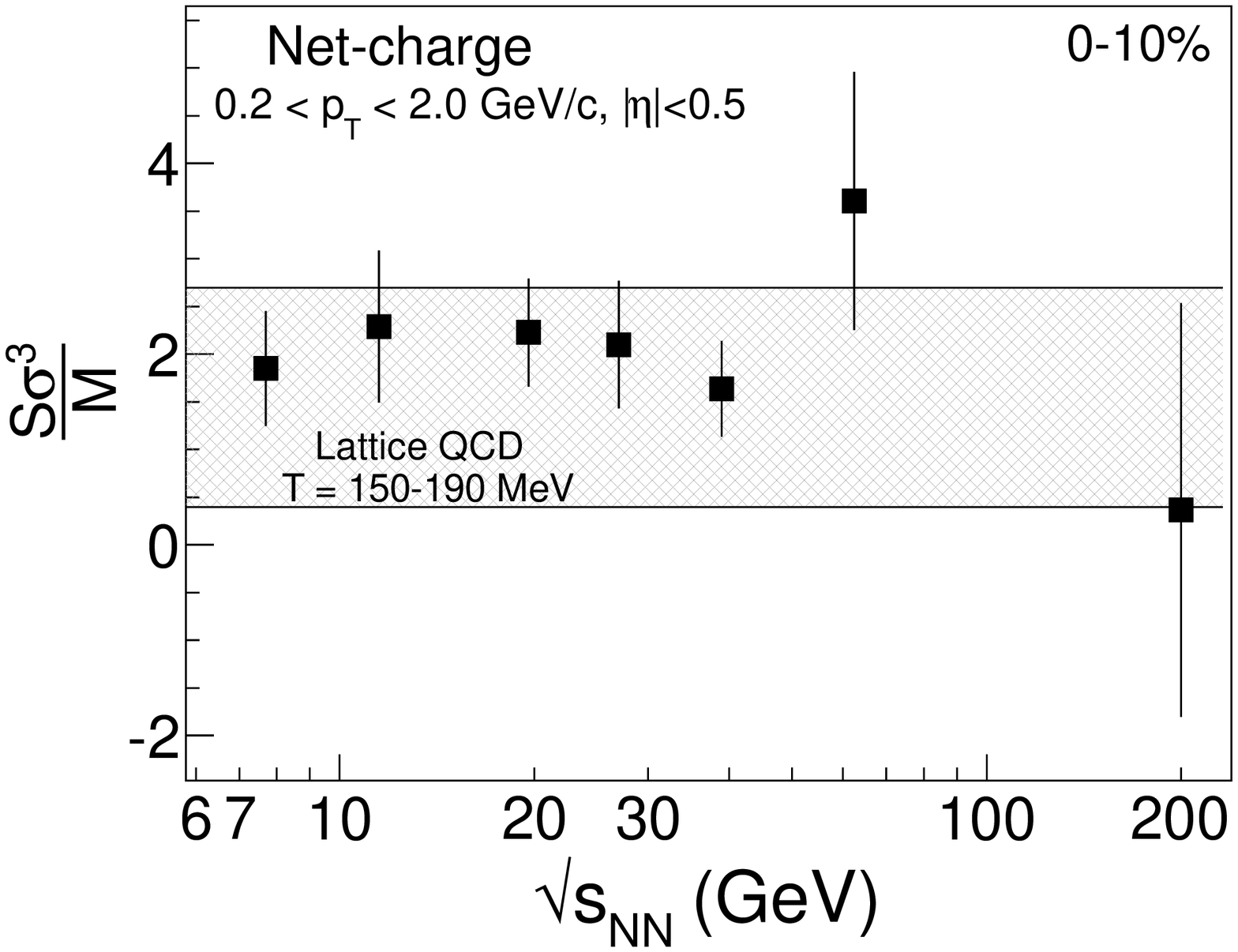}
\end{center}
 
\caption{ \label{FO_netQ}  The beam energy dependance of efficiency
  corrected $\frac{M}{\sigma^{2}}$ (left) $\frac{S\sigma^{3}}{M}$
(right) are shown at $0-10\%$ centrality for Au+Au collisions~\cite{Nihar_thesis}. The
shaded region in right side plot is from Lattice QCD estimation~\cite{Bazavov_Borsanyi_FO}.}
\end{figure}

Besides being a signature of the critical point, the experimentally measured higher-moments of
net-charge and net-proton distributions provide the information to
extract freeze-out conditions in heavy-ion-collisions using lattice QCD
calculation~\cite{Bazavov_Borsanyi_FO}. In Fig~\ref{FO_netQ}, $\frac{M}{\sigma^{2}}$  ({\it
  Baryometer}) decreases as a function collision energy, which implies with the increase 
in colliding energy baryon chemical potential decreases at freeze-out
surface. The values of $\frac{S\sigma^{3}}{M}$ are close to 2.0 for
\sNN~= 7.7 to 62.4 GeV within statistical uncertainty. Due to the large
statistical uncertainty, ${\mu_{B}}_{f}$ and $T_{f}$ in
heavy-ion-collisions can not be precisely constrained with this
method.  

\subsection{\bf {Transverse momentum fluctuations}}
Transverse momentum ($p_{t}$) fluctuations could be envisioned as a source of
temperature fluctuations of the bulk properties of
the system~\cite{Stodolsky}. Two-particle $p_{t}$ correlation scaled by the
ensemble average of $<p_{t}>$, $\sqrt{<\Delta p_{t,i},\Delta
  p_{t,j}>}/<<p_{t}>>$, is also related to the specific heat, $C_{V}$, of the system. Hence, $p_{t}$
fluctuations as a function of \sNN~may provide information about
the order of QCD phase transition. The beam energy dependence of
this observable has been studied for the BES program at mid-rapidity $|\eta|<0.5$~\cite{pTrivedy_QM}.

In Fig.~\ref{pt_ratio_fluct_BES}, the left panel shows the STAR
measurement of this fluctuation observable significantly decreases below
\sNN~= 19.6 GeV for Au+Au collisions at $0-5\%$ centrality, whereas it
remains flat for higher energies (upto ALICE energy). The effects from jet, resonance
decay and finite acceptance effect may contribute to this
observable. Further study is required to draw conclusions about the
phase transition on it.

\subsection{\bf{Particle ratio fluctuations}}
The non-monotonic behavior of the $K/\pi$ yield ratio at \sNN$\sim$7.6
GeV, in SPS energy range for central Pb+Pb collisions, appears to be a
unique characteristic of heavy-ion-collisions~\cite{NA49_Kpi}. This observation
is speculated to be a signature of the phase transition from hadronic to QGP
state. To understand this behavior as a function of collision energy,
dynamical fluctuations of $K/\pi$ ratio and other particle's yield
ratios (like $K/p$, $p/\pi$) have been proposed~\cite{STAR_dyn,NA49_dyn}. STAR collaboration
performed similar analyses in the BES program~\cite{pTrivedy_QM}. 

\begin{figure}
\begin{center}
\includegraphics[width=0.41\textwidth]{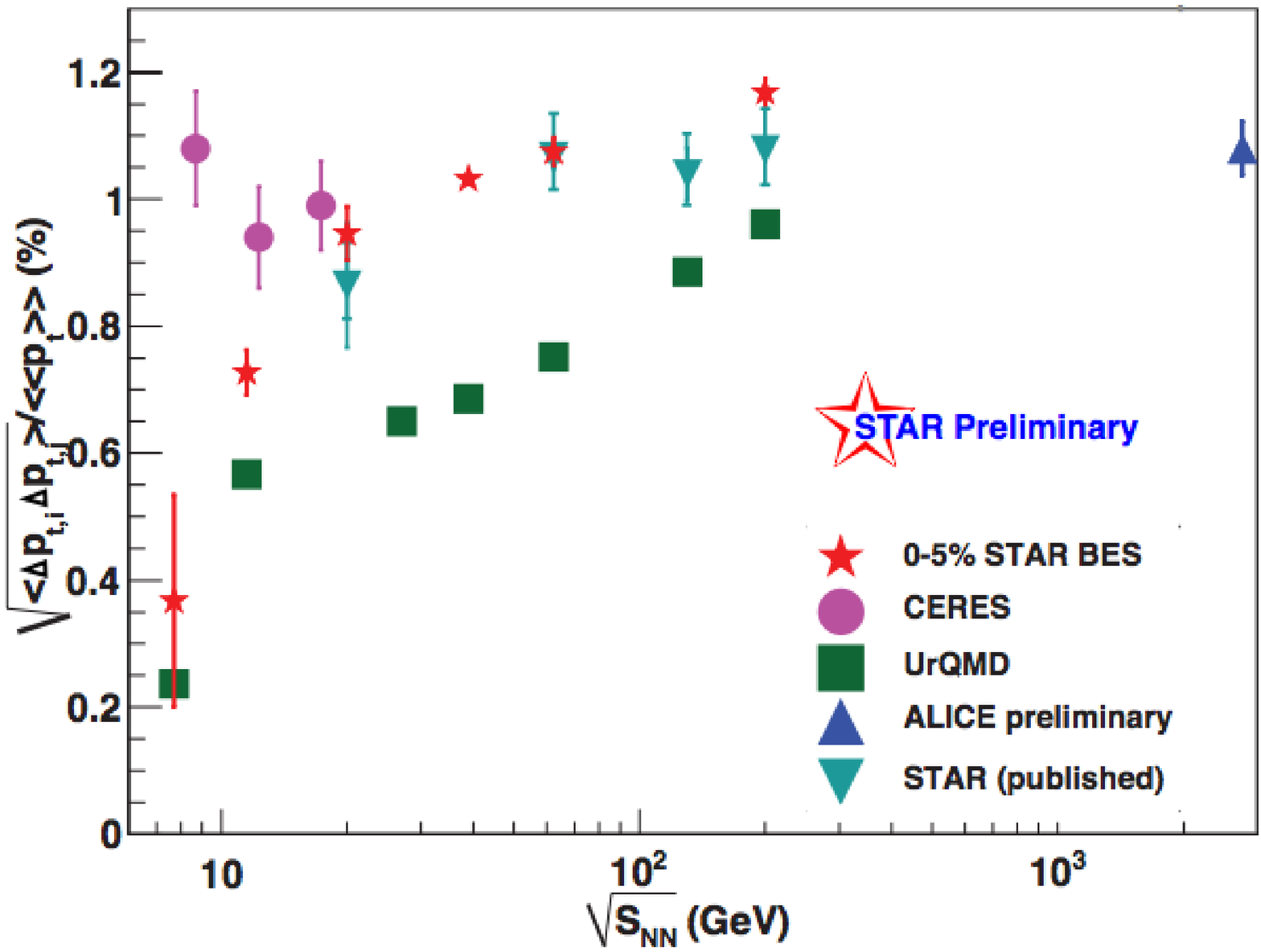}
\includegraphics[width=0.39\textwidth]{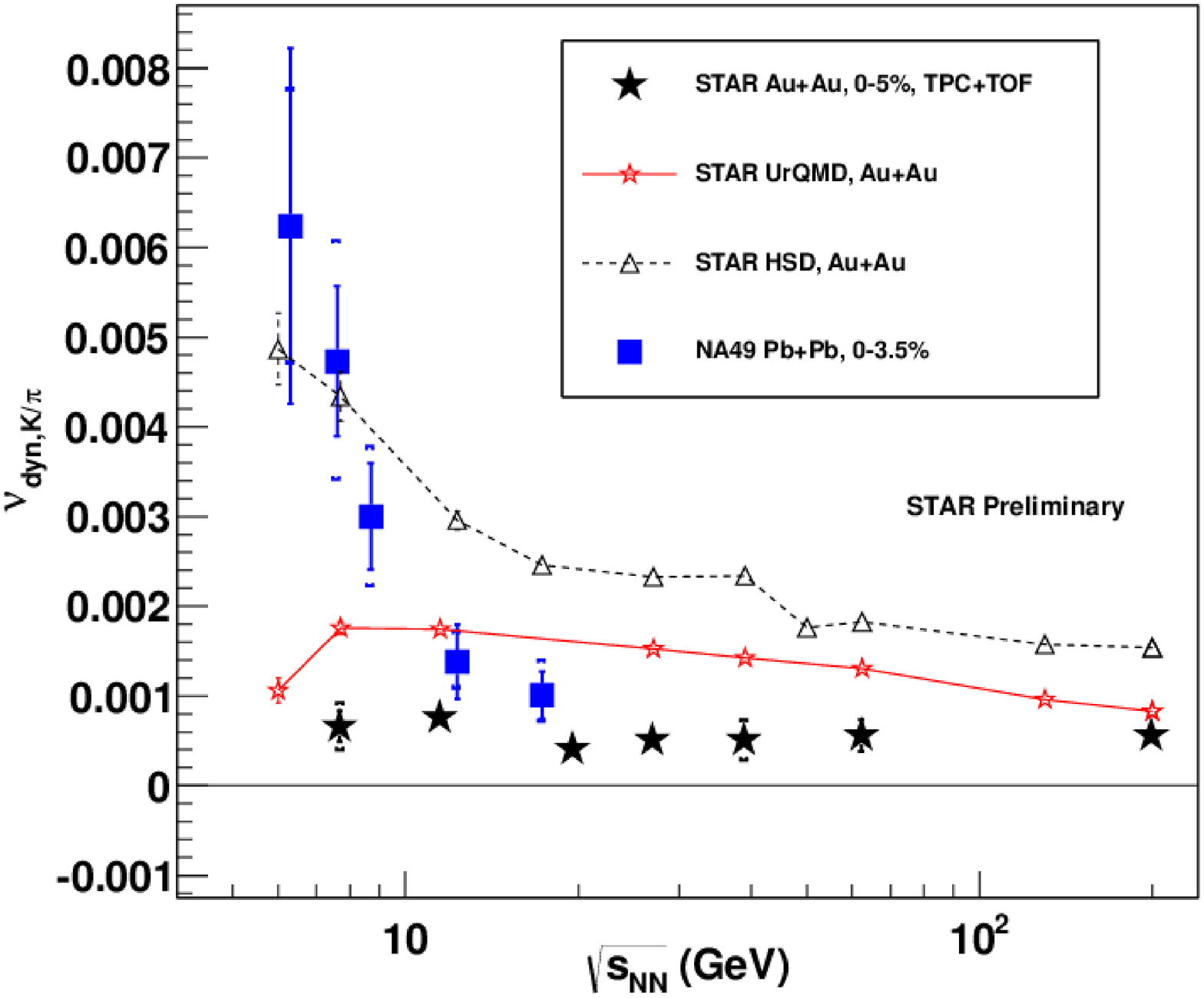} 
\end{center}
 
\caption{ \label{pt_ratio_fluct_BES} (Left) $\sqrt{<\Delta p_{t,i},\Delta p_{t,j}>}/<<p_{t}>>$ as a
  function of \sNN~at $0-5\%$ centrality. (Right) Dynamical fluctuation of kaon to pion ratio,
  $\nu_{dyn,k/\pi}$, as a function of \sNN~at $0-5\%$ centrality~\cite{pTrivedy_QM}.}
\end{figure}

In Fig~\ref{pt_ratio_fluct_BES}, the right panel shows the measure of dynamical
fluctuation $K/\pi$ yield ratio at BES energies at mid-rapidity. This results shows
monotonic behavior as a function of \sNN~at $0-5\%$ centrality for
Au+Au collisions. Similarly, no non-monotonic behavior is observed in case of $K/p$
fluctuations.  It is important to understand the sensitivity of
these observables towards critical fluctuations. A further
understanding of the various mechanisms of
particle production, charge dependence of different particles
species, resonance decays and re-scattering effects on these
observables is necessary.

\begin{figure}
\begin{center}
\includegraphics[width=0.41\textwidth]{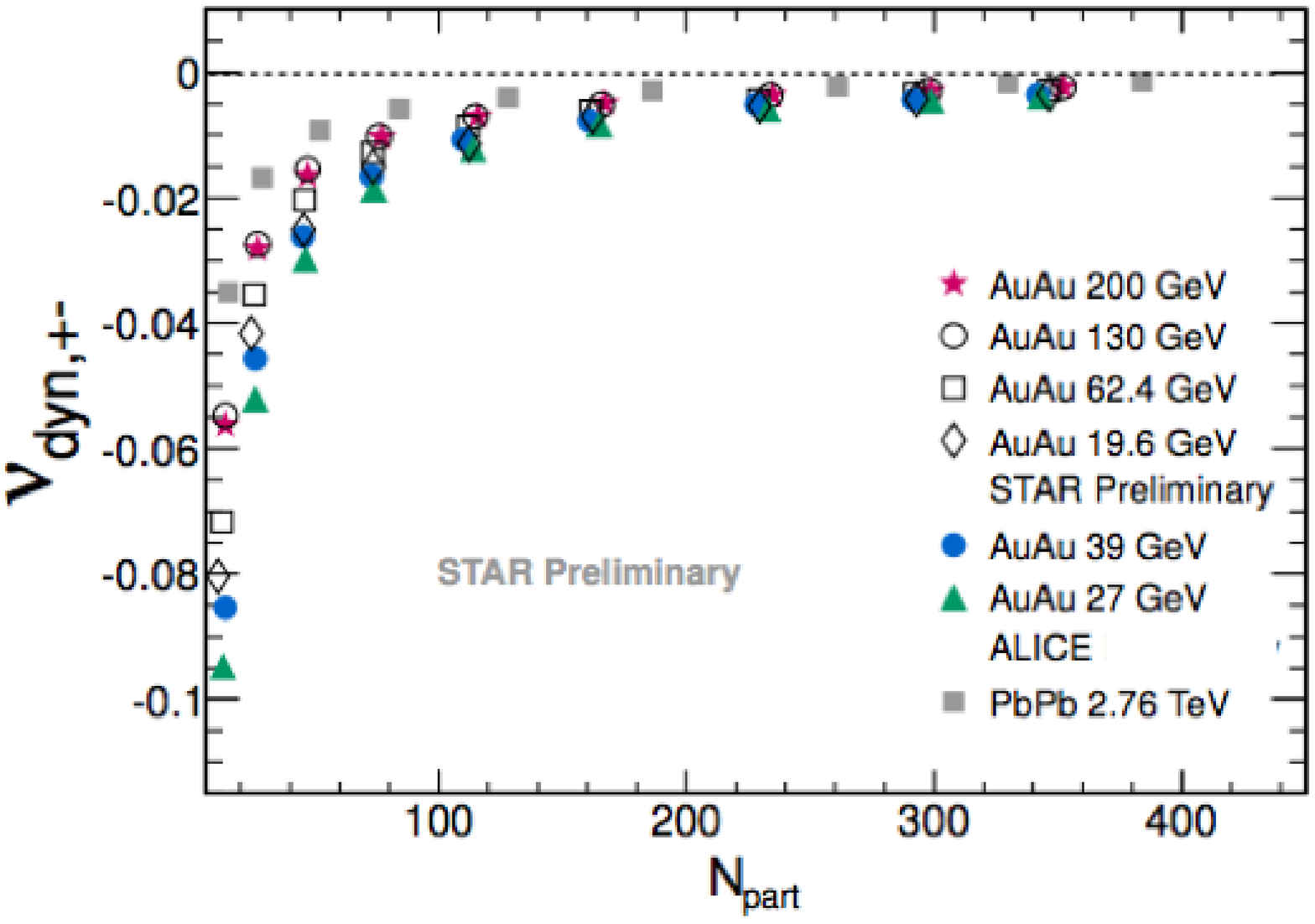}
\includegraphics[width=0.39\textwidth]{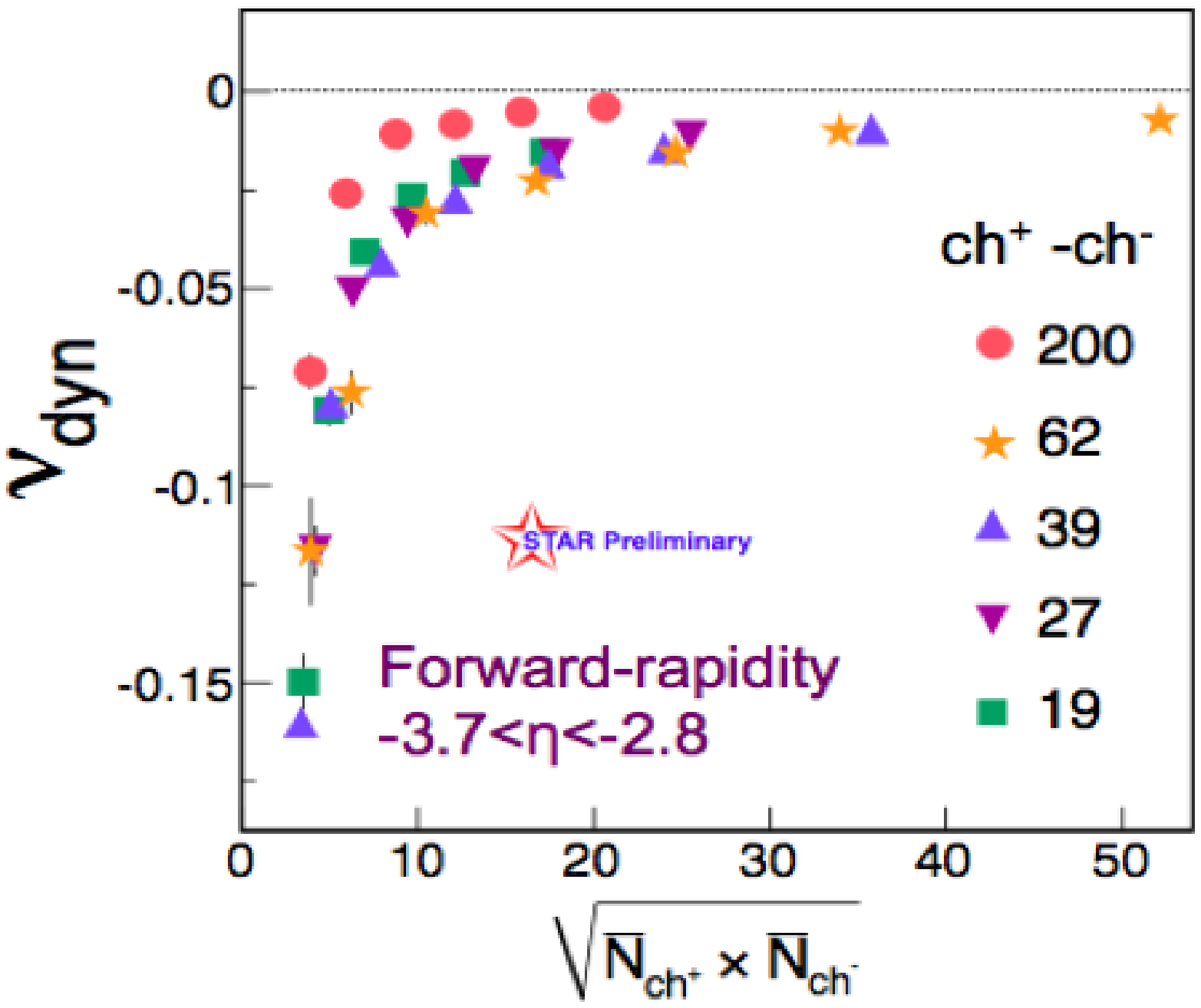} 
\end{center}
 
\caption{ \label{netQ_Frwd_BES} Dynamical Charge fluctuation as
a function of centralities at mid-rapidity (left panel) and Forward
rapidity (right panel)~\cite{pTrivedy_QM}.}
\end{figure}

\section{Dynamical charge fluctuations}
The significant net-charge fluctuations may occur when the QGP phase (state comprises fractional
charge carriers) transits to the Hadron Gas phase (state where the charges are
integral)~\cite{Dyn_Q_fluct}. Such charge fluctuations can be detected by dynamical
charge fluctuations if it could survive the process of evolution of
the system in heavy-ion-collisions. In order to probe such a signature,
$\nu_{dyn}$ for positively and negatively charged particles at mid-rapidity
($|\eta|<0.5$) and forward-rapidity ($-3.7 < \eta < -2.8$) have been
measured~\cite{pTrivedy_QM}, as shown in Fig~\ref{netQ_Frwd_BES}. It is observed that, both
at mid and forward-rapidity,  $\nu_{dyn}$  decreases from central to
peripheral collisions. This observable increases with increasing
collision energies.  This observation may imply the correlation
between positively and negatively charged particles increases with decreasing beam energy or multiplicity. A further study
as a function of different rapidity-windows may help to understand the
diffusion of charge fluctuation in
heavy-ion-collisions~\cite{ALICE_nudy}. Detailed 
study is ongoing to understand more about this observable. 

%\section{Observables related to DCC search}
\section{Future perspective for the BES-II program and Upgrades}
One of the main goals of the BES-I program is to search for the QCD critical
point from event-by-event fluctuation. 
In order to draw a definite conclusion for the location of the QCD critical
point, observables like \KV~and \Ss~of net-charge and net-proton
distributions need a large collection of data. Besides, limited
acceptance in pseudo-rapidity and the centrality determination in STAR
experiment constrain further study of these event-by-event observables. 

The results from BES-I program corroborate for high statistics and
additional beam energies along with detector upgrades for final
conclusion for the QCD critical point. Phase-II of BES program has
been proposed for the years 2018-2019 from \sNN~= 7.7-20 GeV. 
For better particle momentum resolution, dE/dx resolution and improved
pseudo-rapidity acceptance ($\eta|<1.7$), the upgrade of inner sectors
of the current STAR TPC (iTPC) is proposed. An Event Plane and centrality Detector
(EPD) is proposed for the dedicated measurement of event-plane and
centrality determination at forward region $2 < |\eta| < 4$. To achieve
large statistics at low energy, a significant increase of the current
luminosity is planned through the electron cooling.

\section{Summary}
The STAR experiment has successfully collected and analyzed large
collection of data from \sNN~= 7.7 to 39 GeV for its first phase of the
BES program.  The \KV~and \Ss~values of net-proton distribution show significant
deviation from Poisson expectation and Hadron Resonance Gas model
prediction at ~\sNN~= 19.6 and 27 GeV, with large statistical uncertainty at low beam energies. On the other
hand, those of net-charge distribution show no non-monotonic behavior
within large statistical uncertainty. Besides, dynamical fluctuation
of  $K/p, K/\pi$ and $p/\pi$ show monotonicity as a function of beam
energy. The two-particle $p_{t}$ correlation scaled with
average $<p_{t}>$ fluctuation significantly decreases below
\sNN~= 19.6 GeV for Au+Au collisions at $0-5\%$ centrality, whereas it
remains flat for higher energies. Limited statistics and detector acceptance in
BES-I program constrain the ability to draw final conclusion for the exact location and/or
existence of the QCD critical point. The proposed BES-II program with
STAR detector upgrades, like iTPC and EPD, is necessary for further
understanding and drawing final conclusion on the QCD critical point and phase diagram.

\section*{Acknowledgments}

This conference proceedings is supported by the US Department of
Energy under the grant DE-FG02-07ER41485.


\section*{References}
\begin{thebibliography}{9}
\bibitem{Heiselberg} Henning Heiselberg, Physics Reports {\bf 351},
  161 (2001).

\bibitem{Byam}G. Baym, B. Blattel, L.L. Frankfurt, H. Heiselberg,
  M. Strikman, Phys. Rev. C 52 (1995) 1604.

\bibitem{Stodolsky} L. Stodolsky, Phys. Rev. Lett. {\bf 75} (1995).

\bibitem{Stephanov-Rajagopal-Shuryak} M. A. Stephanov, K. Rajagopal
  and E. V. Shuryak, Phys. Rev. Lett. {\bf 81} (1998) 4816; 
  M. A. Stephanov, Int. J. Mod. Phys. {\bf A}, 20, 4387 (2005)
  [arXiv:hep-ph/0402115].

\bibitem{Cleymans} J. Cleymans, H. Oeschler, K. Redlich and
  S. Wheaton, Phys. Rev. {\bf C 73} (2006) 034905.

\bibitem{aoki} Y. Aoki et al., Nature {\bf 443}, 675 (2006).

\bibitem{ejiri} S. Ejiri, Phys. Rev. {\bf D 78}, 074507 (2008).

\bibitem{bowman} E. S. Bowman and J. I. Kapusta, Phys. Rev. {\bf D 79}, 015202 (2009). 

\bibitem{stephanov} M. A. Stephanov,
Prog. Theor. Phys. Suppl. {\bf 153}, 139 (2004); Int. J. Mod. Phys. {\bf A 20}, 4387 (2005).

\bibitem{fodor} Z. Fodor et al., JHEP {\bf 0404}, 50 (2004).

\bibitem{gavai} R. V. Gavai, S. Gupta, Phys. Rev. {\bf D 78}, 114503 (2008).

\bibitem{cheng} M. Cheng et al., arXiv:0811.1006 [hep-lat].

\bibitem{indication} B. Mohanty, J. Alam, S. Sarkar, T.K. Nayak and
  B.K. Nandi, Phys. Rev. {\bf C 68}, 021901 (2003).

\bibitem{berdnikov}
B. Berdnikov and K. Rajagopal, Phys. Rev. {\bf D 61}, 105017 (2000).

\bibitem{athanasiou}
C. Athanasiou, K. Rajagopal,and M. Stephanov, Phys. Rev. {\bf D 82},
074008 (2010).

\bibitem{SG_PLB} R. V. Gavai, S. Gupta, Phys. Lett. {\bf B 696} 459,
 (2011).

\bibitem{Stephanov_corrl}M. A. Stephanov, Phys. Rev. Lett. {\bf 102}
  032301 (2009).

\bibitem{star_net_charge} L. Adamczyk {\it et al.} (STAR
  Collaboration), arXiv:1402.1558, [nucl-ex].

\bibitem{star_net_proton}
L. Adamczyk {\it et al.} (STAR Collaboration), Phys. Rev. Lett. {\bf
  112} 032302 (2014).
 
\bibitem{CBW} X. Luo, J. Xu, B. Mohanty and N. Xu, Jour. Phys. G:
  Nucl. Part. Phys. {\bf 40}, 105104 (2013); N. R. Sahoo, S. De and T. K. Nayak, Phys. Rev. {\bf C 87}, 044906 (2013).

\bibitem{Bzdak}A. Bzdak and V. Koch, Phys. Rev. C 86, 044904 (2012).

\bibitem{STAR_netQ_prelim}
N. R. Sahoo ( for the STAR Collaboration), Acta
Phys. Polon. Supp. 6,  437 (2013); 	arXiv:1212.3892 [nucl-ex];
D. McDonald (for the STAR Collaboration), arXiv:1210.7023 [nucl-ex].

\bibitem{Nihar_thesis}
N. R. Sahoo, Ph. D. thesis, Homi Bhabha National Institute, 2013.

%\bibitem{Bazavov_FO} A. Bazavov et al., Phys. Rev. Lett. {\bf 109}
%  192302 (2012).

\bibitem{Bazavov_Borsanyi_FO} A. Bazavov et al., Phys. Rev. Lett. {\bf 109}
  192302 (2012); S. Borsanyi et al., Phys. Rev. Lett. {\bf 111} 062005 (2013).

\bibitem{NA49_Kpi} S. V. Afanasiev et al. (The NA49 Collaboration),
  Phys. Rev. {\bf C 66} 054902 (2002).

\bibitem{STAR_dyn}
B. I. Abelev {\it et al.} (STAR Collaboration), Phys. Rev. Lett. {\bf
  103} 092301 (2009).

\bibitem{NA49_dyn}
C. Alt  {\it et al.} (NA49 Collaboration), Phys. Rev. {\bf C 79},
044910 (2009).

\bibitem{Dyn_Q_fluct}
S. Jeon and V. Koch, Phys. Rev. Lett. {\bf 85}, 2076 (2000); Masayuki Asakawa, Ulrich Heinz, and Berndt Muller,
Phys. Rev. Lett. {\bf 85}, 2072 (2000).

\bibitem{pTrivedy_QM}
P. Tribedy [STAR Collaboration], Nucl. Phys. A 904-905, 463c (2013);
[arXiv:1211.0171 [nucl-ex]].

\bibitem{ALICE_nudy}
B. Abelev et al. (ALICE Collaboration), Phys. Rev. Lett. {\bf 110}, 152301 (2013).

%\bibitem{STAR_BES_WP}
%STAR BES-I White Paper,{\it ``Studying the Phase Diagram of QCD Matter at RHIC''}. 

\end{thebibliography}
\end{document}